\documentclass[prd,preprint,tightenlines,showpacs,amsmath,amssymb]{revtex4}

\usepackage{epsfig,epsf,graphics,psfrag}
\usepackage{times}
\usepackage{float}
\usepackage{color}
\usepackage{amsfonts,amssymb,stmaryrd,latexsym,amsmath}
\usepackage{multirow}
\usepackage{array} 
\usepackage{enumerate}
\usepackage{graphicx}
\usepackage{keyval,graphicx}

\begin{document}

\bibliographystyle{unsrt}

\title{Searching for observable effects induced by anomalous triangle singularities}

\author{Xiao-Hai Liu$^1$\footnote{liuxh@th.phys.titech.ac.jp}, Makoto Oka$^{1, 2}$\footnote{oka@th.phys.titech.ac.jp} and Qiang Zhao$^3$\footnote{zhaoq@ihep.ac.cn}}

\affiliation{$^1$ Department of Physics, H-27, Tokyo Institute of Technology, Meguro, Tokyo 152-8551, Japan}

\affiliation{$^2$ Advanced Science Research Center, JAEA, Tokai, Ibaraki 319-1195, Japan}

\affiliation{$^3$ Institute of High Energy Physics and Theoretical Physics Center for Science Facilities,
        Chinese Academy of Sciences, Beijing 100049, China}

\date{\today}

\begin{abstract}

We investigate the anomalous triangle singularity (ATS) and its possible manifestations in various processes. We show that the ATS should have important impact on our understanding of the nature of some newly observed threshold states. Discussions on how to distinguish the ATS phenomena from genuine dynamic pole structures are presented. \\

\end{abstract}

\maketitle

\section{Introduction}

The newly observed resonance-like structures in high-energy experiment have intrigued a lot of studies of hadron spectroscopy. These structures are notated as ``$XYZ$" states among which some cannot be accommodated in the conventional quark model. It has been a critical issue whether these resonance-like structures are genuine particles, such as multi-quark states, hybrids, or molecular states, or simply kinematic effects. For the latter ones, we specifically refer to the threshold CUSP effects produced by the two-body branch points proposed in the literature. In a recent work, it is demonstrated that the $S$-wave threshold enhancement could be related to a pole structure if such a threshold enhancement also appears predominantly in its elastic channel~\cite{Guo:2014iya}. The analysis of Ref.~\cite{Guo:2014iya} provides a possible method for distinguishing a genuine pole from the CUSP effects. Meanwhile, it is also pointed out that the kinematic singularity, namely, the so-called ``anomalous triangle singularity (ATS)", if located at specific kinematic region, can produce resonance-like structures.

The possible manifestation of the ATS of the $S$-matrix elements was first noticed in 1960s and theoretical attempts were made to try to clarify the resonance-like structure, i.e. whether they are caused by the ATS or they are genuine resonance peaks~\cite{Peierls:1961zz,Goebel:1964zz,hwa:1963aa,landshoff:1962aa,Aitchison:1969tq}. These theories are based on the study of the analytic properties of the $S$-matrix. It was pointed out that in certain circumstances the ATS of the scattering amplitude may result in observable resonance-like structures when the singularities approach the physical region. Unfortunately, most of those proposed cases were lack of experimental support and our knowledge about how such a kinematic singularity manifests itself was still limited.

In 2012 the BESIII Collaboration published their measurement of the radiative decay $J/\psi\to\gamma +\eta(1405/1475)$ in the exclusive decay channel of $\eta(1405/1475)\to f_0(980)\pi\to  3\pi$~\cite{BESIII:2012aa}. It was found that the isospin-violating decay of $\eta(1405/1475)$ was anomalously large and could not be explained by the $a_0(980)-f_0(980)$ mixing. Theoretical interpretation was provided in Ref.~\cite{Wu:2011yx} where it was proposed that the triangle singularity plays a crucial role to enhance the isospin-violating effects. It can be examined easily that the kinematical condition for the ATS is perfectly satisfied in this process and the signature was stamped by the narrow peak of the $f_0(980)$ which is generated by the charged and neutral $K\bar{K}$ thresholds. A later detailed analysis suggests that the BESIII data for the enhancement of $\eta(1405/1475)$ may contain a small contribution from $f_1(1420)$ in the $3\pi$ decay channel which can be disentangled by the angular distributions of the pion and the recoiled photon~\cite{Wu:2012pg}. Similar analysis can be found in Ref.~\cite{Aceti:2012dj} where the two-body $\pi\pi$ final state interaction was considered. It is worth mentioning that the $\eta(1405/1475)$ decay through the $K\bar{K}^*(K)$ loop is the first clear manifestation of the triangle singularity in a physical process. A confirmation of this scenario is the observation of signals of $a_1(1420)$ at COMPASS in $\pi^- p \to a_1(1420)^\pm\pi^\mp n\to \pi^+\pi^-\pi^0 n$. This is the isospin-1 channel for the same ATS mechanism which can be recognized via $a_1(1420)\to K\bar{K}^*(K)\to f_0(980)\pi$~\footnote{This mechanism was first pointed out by Q.Z. at Hadron 2013 in Nara. A detailed analysis following this idea was presented by Ketzer {\it et al.} in Ref.~\cite{Ketzer:2015tqa} and by Wu {\it et al.} in a forthcoming analysis~\cite{wu-etal}.}.

The recent observation of the charged charmonium-like state $Z_c(3900)$ in $e^+e^-\to Y(4260)\to J/\psi\pi\pi$ at BESIII~\cite{Ablikim:2013mio,Ablikim:2013emm} and Belle~\cite{Liu:2013dau} also provide another example for the ATS mechanism to be recognized in physical processes. As studied in Refs.~\cite{Wang:2013cya,Liu:2013vfa}, the first open charm $S$-wave threshold $D\bar{D}_1(2420)+c.c.$ is located at the mass region of $Y(4260)$. Knowing that the $D_1(2420)$ dominantly decays into $D^*\pi$, we find that the $D\bar{D}_1(D^*)$ loop approaches the ATS kinematics and favors the production of the $Z_c(3900)$. Similar situation also applies to $Y(4260)\to Z_c(4020)\pi\to h_c\pi\pi$ where the transition amplitude is enhanced since the ATS is close to the physic kinematical region. In Ref.~\cite{Pakhlov:2014qva} the proposed mechanism for the understanding of the $Z(4430)$ in $B$ decay is actually another recognition of the ATS in the physical region.

With the availability of high precision data from experiment and a lot of observations of threshold structures it motivates us to make a systematic study of the ATS in various processes. This will be essential for our understanding of the nature of some of those resonance-like threshold structures and meanwhile allow us to probe the kinematic singularities in physical processes.

This work is organized as follows: In Sec. II we present a general analysis of the ATS. In Sec. III we discuss physical processes where the singularities are located in the physical kinematic region, thus, could manifest themselves with measurable effects in experiment. A brief summary is given in Sec. IV.

\section{Anomalous triangle singularity}

\begin{figure}[t]
	\centering
	\includegraphics[width=0.27\hsize]{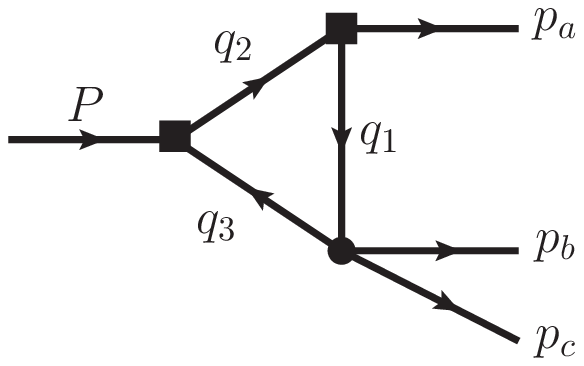}
	\caption{Rescattering process in 3-body decays via the triangle diagram. The internal mass which corresponds to the internal momentum $q_i$ is $m_i$ ($i$$=$1, 2, 3). For the external momenta, we define $P^2=s_1$, $(p_b+p_c)^2=s_2$ and $p_a^2=s_3$.}\label{triangle}
\end{figure}

Kinematic singularities may occur due to the rescattering processes, and the three-body decays are ideal for creating such an environment. We will focus on the triangle diagrams and make an analysis of the analytic properties of the rescattering amplitude.
A typical triangle diagram is illustrated in Fig.~\ref{triangle}. Without losing generality we consider the scalar 3-point function which takes the following form under the Feynman parametrization:
\begin{eqnarray}\label{threepoint}
\Gamma_3(s_1,s_2,s_3)&=&\frac{1}{i(2\pi)^4}\int \frac{d^4 q_1}{(q_1^2-m_1^2+i\epsilon)(q_2^2-m_2^2+i\epsilon)(q_3^2-m_3^2+i\epsilon)}  \nonumber \\
&=& \frac{-1}{16\pi^2}\int_0^1 \int_0^1 \int_0^1 da_1\ da_2\ da_3\ \frac{\delta(1-a_1-a_2-a_3)}{D-i\epsilon} \ ,
\end{eqnarray}
where
\begin{equation}
D \equiv \sum_{i,j=1}^{3} a_i a_j Y_{ij},  \ \
Y_{ij}=\frac{1}{2} \left[ m_i^2+m_j^2-(q_{i}-q_{j})^2 \right] \ .\nonumber
\end{equation}
For this 3-point function $\Gamma_3$, there are several kinds of singularities, and
the location of the singularities in the complex plane of the external momentum variables can be determined by a set of equations, which are usually called the Landau equations~\cite{Landau:1959fi}. In some special kinematic configurations, if all of the three internal lines approach their
on-shell conditions simultaneously, it will correspond to the leading singularity of the triangle diagram~\cite{bonnevay:1961aa} and is what we called the ``ATS". Note that the ATS is different from those singularities in which only two of the internal lines get on shell and such singularities are actually lower-order singularities.  According to the Landau equations, the leading singularity occurs when $\partial D / \partial a_j=0$ is satisfied for all $j$, which will lead to the equation
\begin{eqnarray}\label{landau}
\mbox{det}[Y_{ij}]=0 \ ,
\end{eqnarray}
where $\mbox{det}[Y_{ij}]$ is a function of six variables comprising three external invariant masses $\sqrt{s_i}$ and three internal masses $m_i$ ($i=1,2,3$).
If we fix the internal masses $m_i$, the external invariants $s_1$ and $s_3$, we can obtain the solutions of Eq.~(\ref{landau}) for $s_2$, i.e.,
\begin{eqnarray}
s_2^{\pm}&=&(m_1+m_3)^2+\frac{1}{2m_2^2} {\LARGE[}(m_1^2+m_2^2-s_3)(s_1-m_2^2-m_3^2)-4m_2^2 m_1 m_3 \nonumber \\  &\pm& \lambda^{1/2}(s_1,m_2^2,m_3^2)\lambda^{1/2}(s_3,m_1^2,m_2^2){\LARGE ]},
\end{eqnarray}
with $\lambda(x,y,z)\equiv (x-y-z)^2-4yz$.
Likewise, by fixing $m_i$, $s_2$ and $s_3$ we can obtain the similar solutions for $s_1^\pm$, i.e.,
\begin{eqnarray}
s_1^{\pm}&=&(m_2+m_3)^2+\frac{1}{2m_1^2} {\LARGE[}(m_1^2+m_2^2-s_3)(s_2-m_1^2-m_3^2)-4m_1^2 m_2 m_3 \nonumber \\  &\pm& \lambda^{1/2}(s_2,m_1^2,m_3^2)\lambda^{1/2}(s_3,m_1^2,m_2^2){\LARGE ]}.
\end{eqnarray}
 We will learn later that within the physical boundary only the solution of $s_1^-$ or $s_2^-$ corresponds to the ATS of the amplitude and we call $s_1^-$ and $s_2^-$ as the anomalous thresholds.
For the 3-point function $\Gamma_3$, there is another kind of singularity, i.e. the second-type singularity, which is not associated with the Landau equations~\cite{Eden:1966}. The second-type singularity appears when the three external momenta of the triangle diagram lying along a line, which is irrelevant with the internal masses, and its contribution is not important in the kinematic region that we are interested in. Therefore, we will not discuss it in this work (see Ref.~\cite{Eden:1966} for detailed discussions).

\begin{figure}
	\centering
	\includegraphics[width=0.6\hsize]{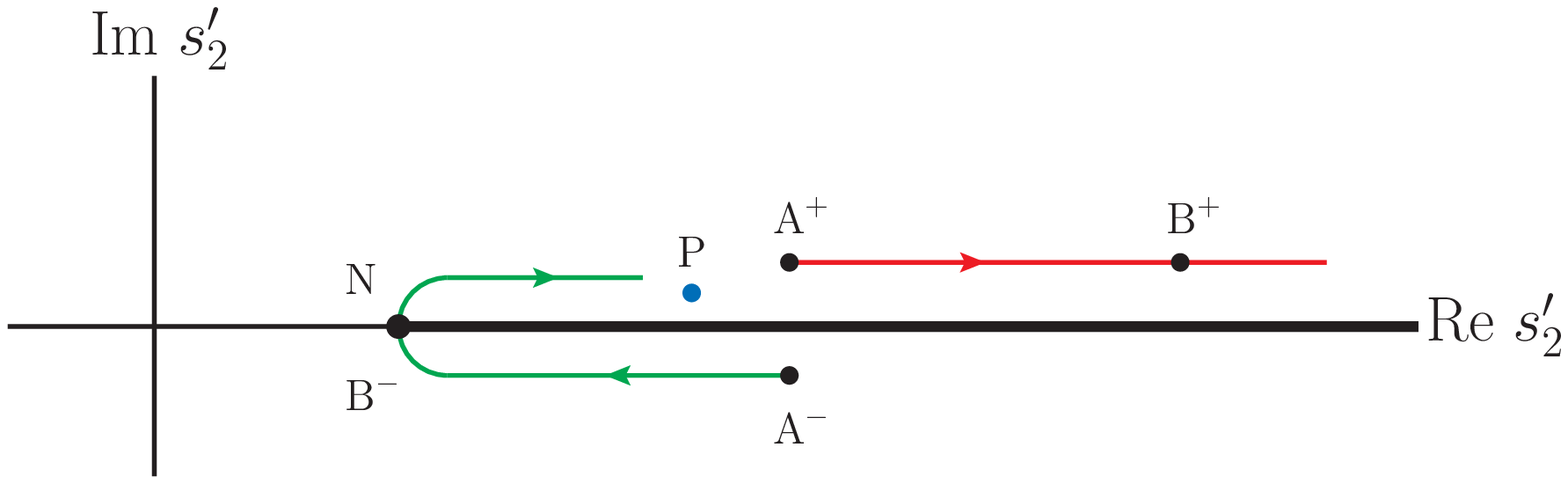}
\caption{Trajectory of $s_2^\pm$ in the complex $s_2^\prime$-plane with $s_1$ increasing from $s_{1N}$ to $\infty$. The thick line on the real axis from the normal threshold point N to $\infty$ is the contour of integration in Eq.~(\ref{singledp}). The $\mbox{A}^+\mbox{B}^+$ ($\mbox{A}^-\mbox{B}^-$) line indicates the trajectory of $s_2^+$ ($s_2^-$). The points are identified as $\mbox{A}^{\pm}$: $s_1=s_{1N}$, $s_2^\pm=s_{2C}\pm i\epsilon$, $\mbox{B}^{-}$: $s_1=s_{1c}$, $s_2^-=s_{2N}$, and $\mbox{B}^{+}$: $s_1=s_{1c}$, $s_2^+=s_{2N}+\frac{m_3}{m_1 m_2^2}\lambda(s_3,m_1^2,m_2^2)+i\epsilon$. The point P indicates another singularity of the integrand in Eq.~(\ref{singledp}), i.e., $s_2+i\epsilon$.}\label{trajectory}
\end{figure}

To elaborate how the ATS occurs, it will be convenient to use the dispersion relation to represent the 3-point function $\Gamma_3$. In the kinematic region $0$$<$$s_1$$<$$(m_2+m_3)^2$, $0$$<$$s_2$$<$$(m_1+m_3)^2$ and $s_3$$<$$(m_2-m_1)^2$, the single dispersion representation of $\Gamma_3$ in $s_2$ takes the form
\begin{eqnarray}\label{singledp}
\Gamma_3(s_1,s_2,s_3)=\frac{1}{\pi}\int\limits_{(m_1+m_3)^2}^{\infty} \frac{ds_2^\prime}{s_2^\prime -s_2 -i\epsilon}\
\sigma(s_1,s_2^\prime,s_3) \ ,
\end{eqnarray}
where the spectral function $\sigma(s_1,s_2,s_3)$ can be obtained by means of the Cutkosky's rules or equally the formula \cite{Cutkosky:1960sp}
\begin{eqnarray}
\sigma(s_1,s_2,s_3) &=& \frac{-1}{16\pi}\int_0^1 \int_0^1 \int_0^1 da_1\ da_2\ da_3\ \delta(1-a_1-a_2-a_3) \delta(D).
\end{eqnarray}
The result reads
\begin{eqnarray}\label{sigma}
\sigma(s_1,s_2,s_3) &=& \sigma_+ - \sigma_-,\nonumber \\
\sigma_{\pm}(s_1,s_2,s_3)&=&\frac{-1}{16\pi \lambda^{1/2}(s_1,s_2,s_3)} \mbox{log}[-s_2(s_1+s_3-s_2+m_1^2+m_3^2-2m_2^2) \nonumber \\
&-& (s_1-s_3)(m_1^2-m_3^2) \pm \lambda^{1/2}(s_1,s_2,s_3)\lambda^{1/2}(s_2,m_1^2,m_3^2)].
\end{eqnarray}
The dispersion representation Eq.~(\ref{singledp}) actually has a larger range of validity \cite{barton:1961aa,fronsdal:1964aa,bronzan:1964aa,norton:1964aa,Lucha:2006vc}. By analytic continuation, it can be extended into the over threshold region
\begin{equation}\label{region}
s_1\geq (m_2+m_3)^2,\ (m_1+m_3)^2\leq s_2 \leq (\sqrt{s_1}-\sqrt{s_3})^2,\ 0\leq \sqrt{s_3}\leq m_2-m_1 \ .
\end{equation}
This specific kinematic region is where the ATS should occur and
it just lie on the physical boundary as part of this region \cite{fronsdal:1964aa,coleman:1965aa}. We will focus the discussion on the kinematic region of Eq.(\ref{region}) in the following sections to demonstrate how the ATS plays a role. For fixed $s_1$, $s_3$ and $m_i$, the spectral function $\sigma(s_1,s_2,s_3)$ has logarithmic branch points $s_2^\pm$, which are just the anomalous thresholds by solving Eq.~(\ref{landau}). We hope to learn how the logarithmic branch points $s_2^\pm$ move as $s_1$ increases from the threshold of $(m_2+m_3)^2$, with $s_3$ and $m_i$  fixed. To obtain the correct analytic continuation of $\Gamma_3(s_1,s_2)$ when $s_1$ exceeds $(m_2+m_3)^2$, it is then necessary to make the substitution $s_1$$\to$$s_1$$+$$i\epsilon$. Thresholds $s_2^\pm$ in the $s^\prime$-plane are then located at
\begin{equation}
s_2^\pm(s_1+i\epsilon)=s_2^\pm(s_1)+i\epsilon \frac{\partial s_2^\pm}{\partial s_1},
\end{equation}
and the corresponding trajectories of $s_2^\pm$ are plotted in Fig.~\ref{trajectory}.
We define the normal thresholds and critical values of the anomalous thresholds  for $s_1$ and $s_2$ as follows,
\begin{eqnarray}\label{s1Ns1C}
&& s_{1N}=(m_2+m_3)^2,\ s_{1C}=(m_2+m_3)^2 +\frac{m_3}{m_1}[(m_2-m_1)^2-s_3], \\
&& s_{2N}=(m_1+m_3)^2,\ s_{2C}=(m_1+m_3)^2 +\frac{m_3}{m_2}[(m_2-m_1)^2-s_3],
\end{eqnarray}
where $s_{1C}$ ($s_{2C}$) is obtained under the condition $\partial s_2^\pm / \partial s_1$$=$$0$ ($\partial s_1^\pm / \partial s_2$$=$$0$).
The imaginary part of $s_2^+$ will always be positive. When $s_1$ increases from $s_{1N}$ to $s_{1C}$, $s_2^-$ moves from $s_{2C}$ (point $\mbox{A}^-$) to $s_{2N}$ (point $\mbox{B}^-$), and lies infinitesimally below the real axis. When $s_1$ exceeds $s_{1C}$, the imaginary part of $s_2^-$ turns to be positive. Therefore, only when $s_{1N}$$\leq$$s_1$$\leq$$s_{1C}$, two singularities of the integrand in Eq.~(\ref{singledp}), i.e. $s_2^-$ and $s_2$$+$$i\epsilon$ (point P), will pinch the contour of integration in the $s_2^\prime$-plane. This pinch singularity which occurs when $s_2=s_2^-$ is a direct manifestation of the ATS of $\Gamma_3$. Likewise, by fixing $m_i$, $s_2$ and $s_3$  we can derive that only when $s_{2N}$$\leq$$s_2$$\leq$$s_{2C}$, the ATS of $\Gamma_3$ will appear at $s_1$$=$$s_1^-$, which lies between $s_{1C}$ and $s_{1N}$.

We define the discrepancy between the normal and anomalous thresholds as follows,
\begin{eqnarray}
	\Delta_{s_1}&=&\sqrt{s_1^-} - \sqrt{s_{1N}}, \nonumber \\
\Delta_{s_2}&=&\sqrt{s_2^-} - \sqrt{s_{2N}}.
\end{eqnarray}
Apparently, when $s_2$$=$$s_{2N}$ ($s_1$$=$$s_{1N}$), we will obtain the maximum value of $\Delta_{s_1}$ ($\Delta_{s_2}$), i.e.,
\begin{eqnarray} \label{deltas1s2}
\Delta_{s_1}^{\mbox{max}}&=&\sqrt{s_{1C}} - \sqrt{s_{1N}}\approx
\frac{m_3}{2m_1(m_2+m_3)}[(m_2-m_1)^2-s_3], \nonumber \\
\Delta_{s_2}^{\mbox{max}}&=&\sqrt{s_{2C}} - \sqrt{s_{2N}}\approx
\frac{m_3}{2m_2(m_1+m_3)}[(m_2-m_1)^2-s_3].
\end{eqnarray}

The difference between the normal and anomalous thresholds are due to the nonvanishing three-vector momenta carried by the rescattering particles. Namely, when those three internal particles approach their on-shell conditions simultaneously, they can still carry nonvanishing three-vector momenta respectively which will contribute to the anomalous threshold in the rescattering.

\section{Physical cases recognizing the ATS}

\subsection{Elastic rescattering processes}

We expect that the ATS may lead to some detectable effects in the rescattering processes when the ATS kinematic conditions are satisfied.

 In Ref.~\cite{schmid:1967aa} it was argued that for the elastic rescattering process, when the corresponding resonance-production tree diagram is added coherently to the triangle rescattering diagram, the effect of the triangle diagram is nothing more than a multiplication of the singularity from the tree diagram by a phase factor. Therefore the singularities of the triangle diagram cannot produce obvious peaks in the total transition rate. This is the so-called Schmid theorem, and we refer to Refs.~\cite{Aitchison:1969tq,Goebel:1982yb,Anisovich:1995ab} for some comments on and further studies of this theorem.
 The expectation of the Schmid theorem needs experimental test and the recent BES-III measurement of $e^+ e^-\to D{\bar D}^*\pi +c.c.$ at the mass of 4.26 GeV turns out to be useful for examining this theorem. As shown by the recent analysis in Ref.~\cite{Guo:2014iya}, the pronounced $D\bar{D}^* +c.c.$ threshold enhancement in the elastic channel of $Y(4260)\to D\bar{D}^*\pi +c.c.$ has a non-perturbative feature in the $D\bar{D}^* +c.c.$ rescattering and should imply the presence of a threshold pole structure. It is interesting to recognize that if the Schmid theorem is correct, the observed threshold enhancement would favor more to be produced by a pole structure in the elastic channel. Otherwise, the ATS will still play a role in association with the pole although it should be emphasized that such an ATS threshold peak should be different from the CUSP effects caused by two-body branch points~\cite{Bugg:2011jr}. Further experimental studies of the energy evolution of the threshold enhancement, e.g. the $Z_c(3900)$, will be useful for disentangling how significant these two contributions are in the production of the threshold enhancement. 

It is important to have a combined analysis of the elastic and inelastic scatterings for processes where the ATS is present. Since the inelastic channel does not have ambiguities from the tree diagram which shares the same on-shell kinematics with the ATS, it is argued that the ATS contribution can be identified more easily in the inelastic channel. We are going to discuss the inelastic rescattering processes in detail later but emphasize that the combined analysis of the elastic and inelastic channels are crucial for understanding the nature of a threshold enhancement~\cite{Guo:2014iya}.

\subsection{Inelastic rescattering processes}

We discuss the inelastic rescattering processes in this subsection. Some examples related to the existing observations are shown in Fig.~\ref{rescatterings} and the detailed discussions are as follows:

\subsubsection{$D_{s1}\to D_s\pi\pi$}
\begin{figure}
	\centering
	\includegraphics[width=0.55\hsize]{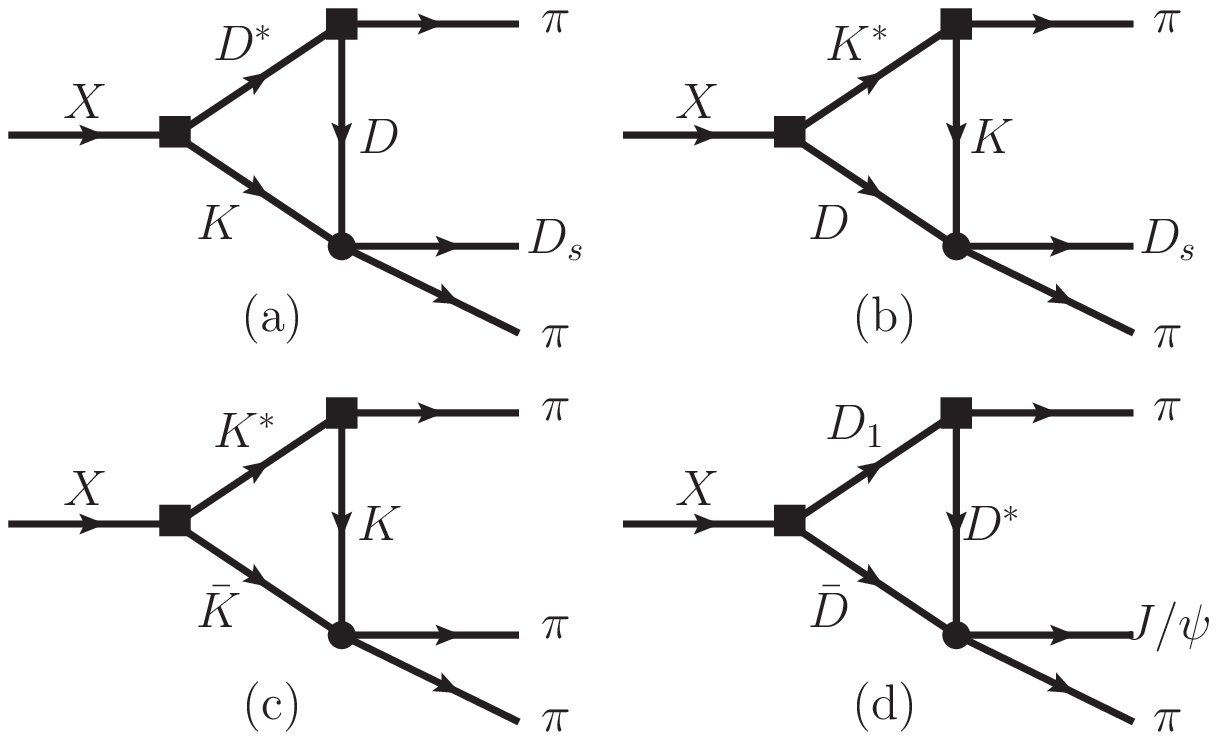}
	\caption{Inelastic rescattering processes in 3-body decays via triangle diagram. $X$ denotes some specified initial state with the proper quantum number, and we define $M_X^2$$=$$s_1$. The conventions of the momenta and invariant masses are the same with those in Fig.~\ref{triangle}.}\label{rescatterings}
\end{figure}

In Fig.~\ref{rescatterings}(a) $X$ stands for the $D_{s1}$ states that couple to $D^* K$ in a relative $S$ wave. In experiment two axial-vector states have been observed, i.e. $D_{s1}(2460)$ and $D_{s1}(2536)$. Various studies have shown that the $D^*K$ open channel has been the most important driving mechanism for shifting the quark model bare states to the physical ones near the $D^*K$ threshold.

\begin{figure} 
	\centering
	\includegraphics[width=0.5\hsize]{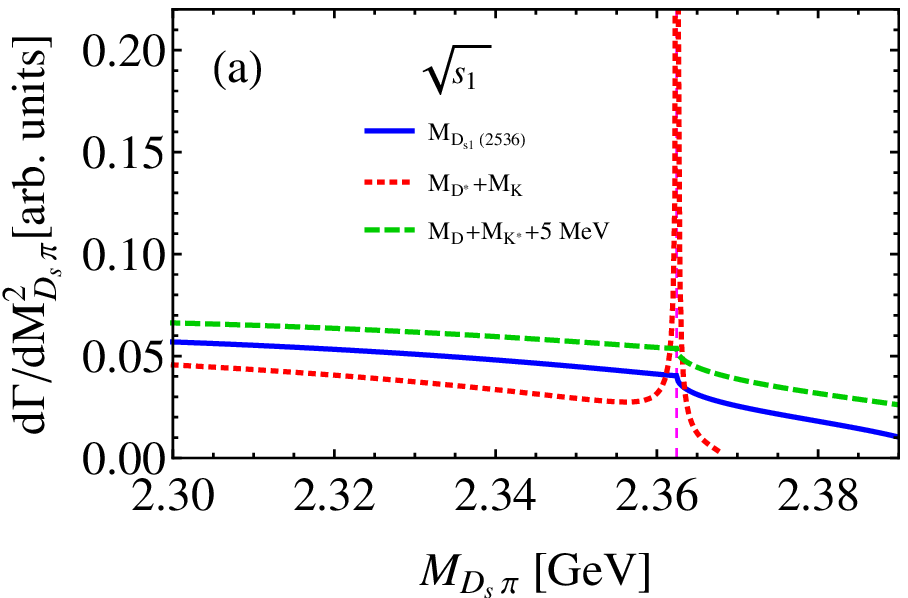}
	\includegraphics[width=0.5\hsize]{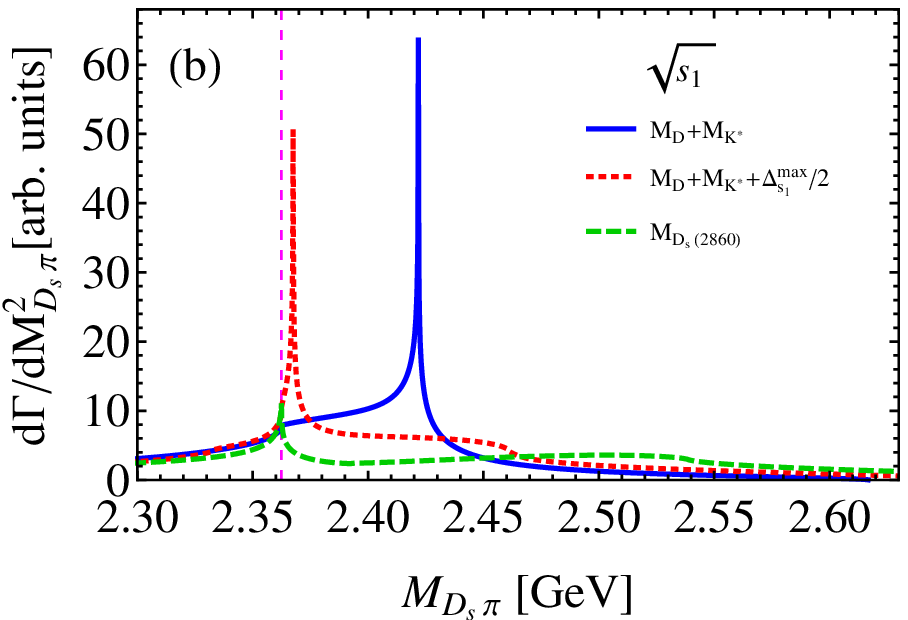}\\
	\includegraphics[width=0.5\hsize]{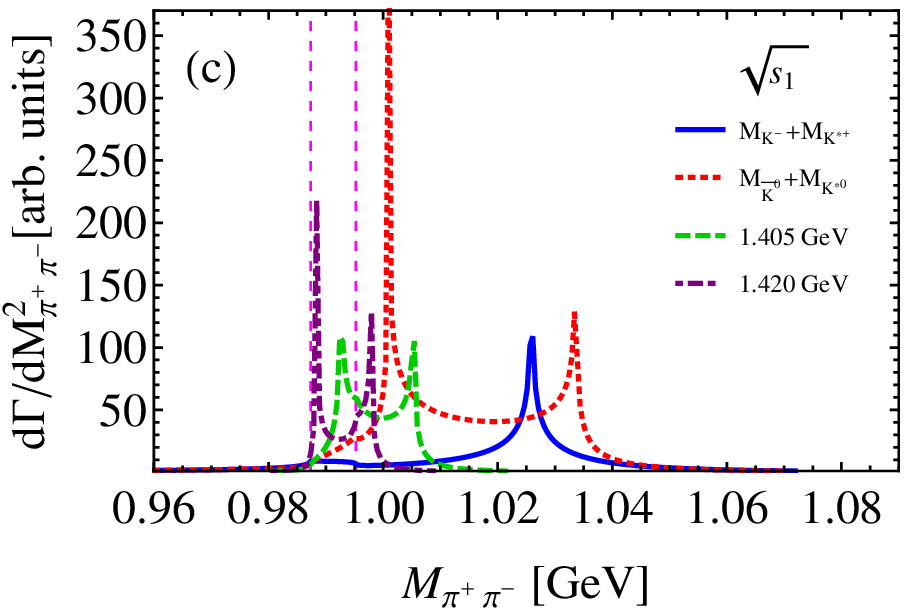}
	\includegraphics[width=0.5\hsize]{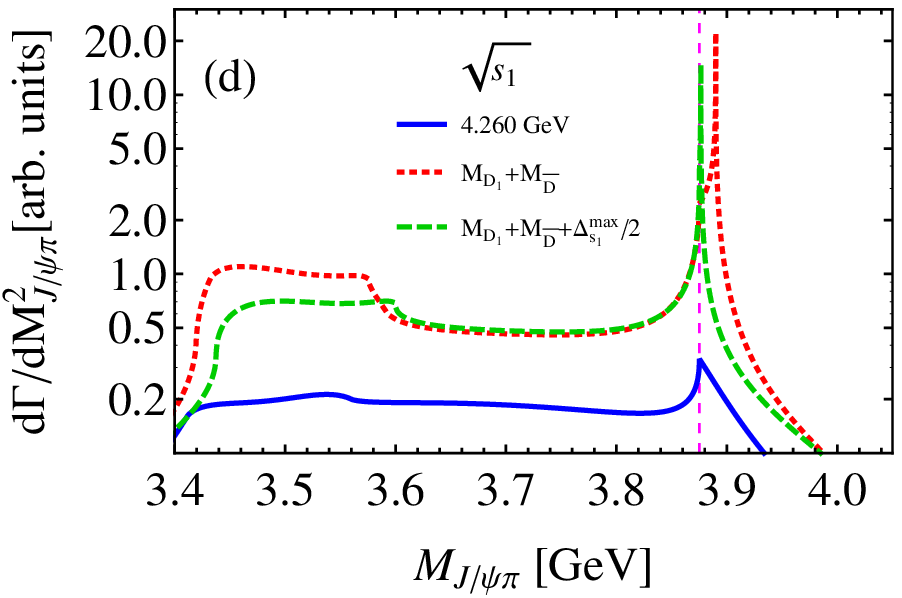}
	\caption{Invariant mass distributions of the corresponding rescattering process in Fig.~\ref{rescatterings}. The vertical dash lines indicate (a): $DK$ threshold, (b): $DK$ threshold, (c): $K^+K^-$ (left) and $K^0\bar{K}^0$ (right) thresholds, and  (d): $D^*\bar{D}$ threshold, respectively. }\label{lineshape}
\end{figure}

The process of Fig.~\ref{rescatterings} (a) satisfies the ATS condition.
We investigate the evolution of the ATS in terms of the initial mass near the $D^*K$ threshold and the corresponding numerical results are displayed in Fig.~\ref{lineshape} (a). In calculating the rescattering amplitudes, we adopt the heavy hadron chiral perturbation theory introduced in~\cite{Casalbuoni:1996pg}. The analytic properties of the rescattering amplitudes mainly depend on the kinematics. Therefore, we only focus on the lineshape behavior of these rescattering processes, but leave the explicit value of the coupling constants to be investigated elsewhere.
As shown by the solid line in Fig.~\ref{lineshape} (a), at the mass of $D_{s1}(2536)$, there is only an unnoticeable cusp appearing at the threshold of $M_D+M_K$. This is because the mass of $D^*$ is very close to the $D\pi$ threshold, which makes the corresponding $\Delta_{s_1}^{\mbox{max}}$ and $\Delta_{s_2}^{\mbox{max}}$ very small, as displayed in Table~\ref{tabdelta}. Only when $M_X$  nearly equals to $M_{D^*} + M_K$, the ATS condition will be fulfilled and there will be a narrow enhancement in the $D_s\pi$ distribution in the vicinity of $DK$ threshold as shown by the dotted line. In fact, only 5 MeV above the $D^*K$ threshold will demolish the ATS enhancement totally as illustrated by the dashed line in Fig.~\ref{lineshape} (a).

\begin{table}
\caption{Kinematic quantities $\Delta_{s_1}^{\mbox{max}}$ and $\Delta_{s_2}^{\mbox{max}}$ for the corresponding Feynman diagram in Fig.~\ref{rescatterings}. Noticed that, when the internal mass $m_i$ ($i=1,2,3$) and $s_3$ are fixed, $\Delta_{s_1}^{\mbox{max}}$ and $\Delta_{s_2}^{\mbox{max}}$ are determined.}\label{tabdelta}
\begin{center}
\begin{tabular}{|c|c|c|c|c|}
	\hline [MeV] & Fig.~\ref{rescatterings}(a) & Fig.~\ref{rescatterings}(b) & Fig.~\ref{rescatterings}(c) & Fig.~\ref{rescatterings}(d) \\
	\hline $\Delta_{s_1}^{\mbox{max}}$ & 0.089 & 96 & 49  & 16 \\
	\hline $\Delta_{s_2}^{\mbox{max}}$ & 0.087  & 62  & 38 & 15  \\
	\hline
\end{tabular}
\end{center}
\end{table}

The above analysis identifies the situation that if one would expect to observe detectable effects caused by the ATS, then $\Delta_{s_1}^{\mbox{max}}$ and $\Delta_{s_2}^{\mbox{max}}$ must be as large as possible. This actually enlarges the kinematic region where the ATS effects can be observable.  According to Eq.~(\ref{deltas1s2}), this requires that the quantity $[(m_2-m_1)^2-s_3]$ should also be as large as possible. Physically, it means that the phase space for a particle with mass $m_2$ decaying into particles with masses $m_1$ and $\sqrt{s_3}$ should be large enough. Taking into account this requirement, one promising rescattering process should be Fig.~\ref{rescatterings} (b). For this triangle diagram,  $\Delta_{s_1}^{\mbox{max}}$ is about 96 MeV and $\Delta_{s_2}^{\mbox{max}}$ is about 62 MeV, which are sizable. This is because the phase space for $K^*$ decaying into $K\pi$ is quite large, and the ratios $M_D/M_{K^*}$ and $M_D/M_{K}$ ($m_3/m_2$ and $m_3/m_1$) are also relatively larger. Then, if $M_X$ is fixed at a value between $M_D+M_{K^*}$ and $M_D+M_{K^*}+\Delta_{s_1}^{\mbox{max}}$, there will be a pronounced peak appearing between $M_D+M_{K}+\Delta_{s_2}^{\mbox{max}}$ and $M_D+M_{K}$ in the invariant mass spectrum of $D_s\pi$.
Both $D_{s1}(2860)$ and $D_{s1}(2700)$ are good candidates for the initial state $X$. These two states are very broad, of which the decay widths are about 159 MeV and 117 MeV, respectively~\cite{Aaij:2014baa,Agashe:2014kda}. Although their pole masses are out of the kinematic region where the ATS can occur, their shoulder or tail can still fall into the kinematic region of the ATS, which may cause some detectable effects.

\begin{figure}[t]
	\centering
	\includegraphics[width=0.5\hsize]{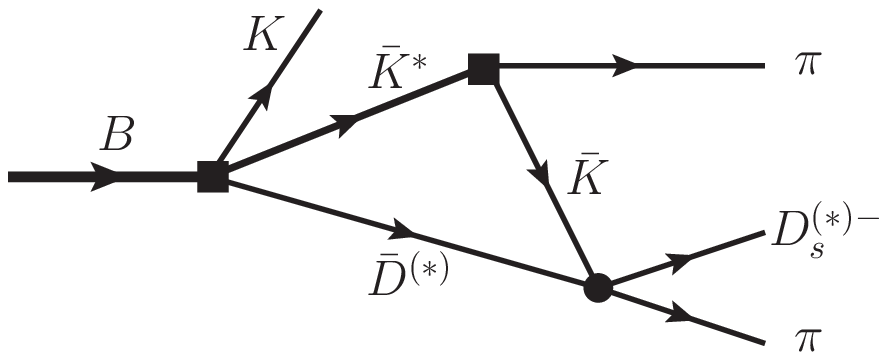}
	\caption{Rescattering process in bottomed meson decays.}\label{Bdecay}
\end{figure}

Similar to Fig.~\ref{rescatterings} (b) the decay of $B$$\to$$K\bar{K}^*\bar{D}^{(*)}$$\to$$KD_s^{(*)-}\pi\pi$ also provides access to the $\bar{K}^*\bar{D}^{(*)}$ rescattering into $D_s^{(*)-}\pi\pi$.  The triangle diagram is illustrated in Fig.~\ref{Bdecay} and the $\bar{K}^*\bar{D}^{(*)}$ rescattering is similar to the process of  Fig.~\ref{rescatterings} (b).
The branching ratio of $B$ decaying into $K\bar{K}^*\bar{D}^{(*)}$ is at the order of $10^{-4}\sim 10^{-3}$, which is sizable~\cite{Agashe:2014kda}. The advantage of this process compared with Fig.~\ref{rescatterings} (b) is that the invariant mass of $D_s^{(*)-}\pi\pi$ can vary in a certain range and one can follow the evolution of the ATS peak continuously.

As discussed for Fig.~\ref{rescatterings} (b), if we fix the invariant mass of $D_s^{(*)-}\pi\pi$ in the energy range from $M_{K^*}+M_{D^{(*)}}$ to $M_{K^*}+M_{D^{(*)}}+\Delta_{s_1}^{\mbox{max}}$, we may find a narrow peak in the $D_s^{(*)-}\pi$ distribution as shown by the curves in Fig.~\ref{lineshape} (b). This resonance-like peak has an exotic flavor quantum number. An interesting feature arising from this energy dependence of the ATS peak evolution is that when the invariant mass of $D_s^{(*)-}\pi\pi$ is fixed at $M_{K^*}+M_{D^{(*)}}$, the location of the peak will be far away from the normal threshold $M_{D^{(*)}}+M_{K}$ as shown by the solid line in Fig.~\ref{lineshape} (b). The energy-dependence of the peak position should offer us a criterion to distinguish an ATS kinematic effect from a pole structure such as a hadronic molecule state.

\subsubsection{$\eta(1405)/\eta(1475)$ and/or $f_1(1420)/a_1(1420)$ decays into $3\pi$}

The processes $\eta(1405)/\eta(1475)$ and/or $f_1(1420)/a_1(1420)$ decaying into $3\pi$ are ideal places satisfying the ATS condition. The corresponding  process is illustrated by Fig.~\ref{rescatterings} (c) where the intermediate $K\bar{K}^*+c.c.$ rescattering by exchanging a $K$ or $\bar{K}$ satisfies the ATS condition perfectly. It should be noted that $\Delta_{s_1}^{\mbox{max}}$ and $\Delta_{s_2}^{\mbox{max}}$ are not small in this case. If we take the initial state $X$ within the range of 1.385$\sim$1.442 GeV, there will be peaks appeared around $K\bar{K}$ threshold in the $\pi^+\pi^-$ invariant mass spectrum as displayed in Fig.~\ref{lineshape} (c). This is the mass region where $\eta(1405)/\eta(1475)$ and $f_1(1420)$ are present and the ATS accounts for their anomalously large isospin violations~\cite{BESIII:2012aa}.

It was first proposed in Ref.~\cite{Wu:2011yx} that the ATS can account for the anomalously large isospin violations for $\eta(1405)/\eta(1475)\to 3\pi$ measured recently by BESIII~\cite{BESIII:2012aa}. Nevertheless, this mechanism will interfere with the tree diagram for $\eta(1405)/\eta(1475)\to \eta\pi\pi$ and result in different peak positions and lineshapes for the initial state which could be either $\eta(1405)$ or $\eta(1475)$. This immediately raises the question whether the experimental observations of two states, $\eta(1405)$ and $\eta(1475)$, in different channels should originate from one single state~\cite{Wu:2011yx}.
In Ref.~\cite{Wu:2012pg}, a detailed analysis of the BESIII data for $J/\psi \to \gamma +3\pi$ suggests that the $f_1(1420)\to 3\pi$ also satisfies the ATS condition and it implies large isospin violations in $f_1(1420)\to 3\pi$ channel.

It is natural and interesting to recognize that the ATS will also give rise to an enhancement around 1.385$\sim$1.442 GeV in the $3\pi$ invariant mass spectrum in the isospin-1 channel. In the $S$-wave the quantum number is $I,J^{PC}=1,1^{++}$ as a partner structure of the $f_1(1420)$, and in the $P$-wave the quantum number is either $I, \ J^{PC}= 1, \ 0^{-+}$ or $0, 1^{--}$ for the neutral states. It should be noted that the recent COMPASS observation of an isovector $a_1(1420)$~\cite{Adolph:2015pws} in $\pi^- p \to a_1(1420)^\pm\pi^\mp n\to \pi^+\pi^-\pi^0 n$ could be a direct recognition of the ATS. A detailed analysis based on the ATS will be presented in Ref.~\cite{wu-etal}.

\subsubsection{$Y(4260)\to J/\psi\pi\pi$}

Another kinematic region which has access to the ATS is the $Y(4260)$ decays into $J/\psi\pi\pi$ if it has a large coupling to $D_1(2420)\bar{D}+c.c.$ This is the process that the charged charmonium-like state $Z_c(3900)$ was observed~\cite{Ablikim:2013mio,Liu:2013dau,Xiao:2013iha}. As pointed out in Ref.~\cite{Wang:2013cya} the $D_1(2420)\bar{D}+c.c.$ threshold is the first $S$-wave open charm threshold with narrow charmed mesons in the vector sector. The closeness of $Y(4260)$ to the $D_1(2420)\bar{D}+c.c.$ threshold makes it a possible candidate for hadronic molecule state of $D_1(2420)\bar{D}+c.c.$ A systematic investigation of such a scenario can be found in Refs.~\cite{Wang:2013hga,Cleven:2013mka,Wang:2013kra,Liu:2013vfa,Liu:2014spa,Wu:2013onz,Szczepaniak:2015eza}.

In the decay of $Y(4260)\to J/\psi\pi\pi$ via the intermediate $D_1(2420)\bar{D}+c.c.$ rescattering the quantities
$\Delta_{s_1}^{\mbox{max}}$ and $\Delta_{s_2}^{\mbox{max}}$ are enlarged due to the large value of $[(m_2-m_1)^2-s_3]$ in the triangle transition displayed by Fig.~\ref{rescatterings} (d).
The process of $D_1(2420)\to D^*\pi$ is the dominant decay channel of the $D_1(2420)$ and satisfies this kinematic requirement.
It is interesting to note that for this kind of charmed meson loops, the normal thresholds are much larger than those corresponding to Figs.~\ref{rescatterings}(b) and (c). As a consequence,  the quantities $\Delta_{s_1}^{\mbox{max}}$ and $\Delta_{s_2}^{\mbox{max}}$ are not very large according to Eq.~(\ref{deltas1s2}). The corresponding values are listed in Table~I.
The ATS peak will then stay close to the normal threshold, as illustrated in Fig.~\ref{lineshape} (d). In this sense, it would be difficult to distinguish the ATS peak from the pole structure in the invariant mass of the $J/\psi\pi$. We shall come back to the relevant issue later in this Section. It should be mentioned that for the solid line in Fig.~\ref{lineshape} (d), although 4.26 GeV is a little bit smaller than the $D_1 D$ threshold, it is still very close to the ATS kinematic region and the physical rescattering amplitude can be enhanced by the singularities to some extent.

Similar to $Y(4260)\to J/\psi\pi\pi$, other kinematics which satisfy the ATS conditions have been explored in both charmonium and bottomonium sectors~\cite{Wang:2013hga,Liu:2013vfa,Liu:2014spa}.

In the above discussions, when the kinematic conditions in Eq.~(\ref{region}) are satisfied, the particle with mass $m_2$ can then decay into two particles with masses $m_1$ and $\sqrt{s_3}$ which allows all the internal particles to approach their on-shell kinematics simultaneously. It should be noted that the width effects from the internal particles will weaken the ATS peak rather apparently~\cite{Wang:2013hga,Liu:2013vfa,Liu:2014spa}. Thus, our discussions on the ATS phenomena are naturally limited to the triangle transitions where only the narrow states are involved, e.g. the widths of $K$ and $D$ are rather small, and the $K^*$, $D^*$ and $D_1$ are also regarded as relatively narrow states. In this paper, we have not taken into account the width in the calculations.

\subsection{ATS peak and pole structure}

There have been a lot of discussions on how to distinguish kinematic effects from a dynamical pole structure in the literature. Here, we would like to first distinguish the kinematic CUSP effects from the ATS effects although both are kinematic effects. As pointed out at the beginning, the CUSP effects are caused by the two-body branch points while the ATS peak is due to more singular conditions required for the triangle transitions. As a consequence, the effects induced by the ATS will be more obvious than that induced by the usual two-body branch points at the normal thresholds.
In Ref.~\cite{Guo:2014iya} a method was developed for distinguishing the pole structures from the kinematic CUSP effects. As emphasized in Ref.~\cite{Guo:2014iya}, a combined measurement of the elastic and inelastic channels for a threshold enhancement would be crucial for disentangling the nature of the threshold enhancement. However, the situation would become complicated if the threshold enhancement also falls into the ATS kinematic region. In such a case, the key question is whether one can distinguish the ATS effects from the dynamic pole structure. Based on what we have learned from the ATS, we propose some criteria that can be implemented into further studies of the threshold states:

i) Since the pole position of a genuine state should not depend on a specific process, while the ATS peak is rather sensitive to the kinematic condition, one would expect that a genuine state should still appear in other processes where the kinematic conditions for the ATS are not fulfilled, but the ATS peak should disappear.

ii) One can investigate different production processes to check how strongly the signal is process-dependent.

\section{Summary}

In this work we made a detailed analysis of the ATS and explored possible channels which allow experimental measurements of this unique mechanism. The ATS can produce observable phenomena which may have important consequences. One example is the puzzling $\eta(1405)$ and $\eta(1475)$ relation. So far, the high-statistic data do not support two states to appear in the same channel. Meanwhile, the single state, either $\eta(1405)$ or $\eta(1475)$, appears to have different mass positions and invariant mass lineshapes in different channels. Such a phenomenon can be naturally explained by the ATS mechanism. Nevertheless, it naturally accounts for the appearance of the $a_1(1420)$ in $\pi^- p \to a_1(1420)^\pm\pi^\mp n\to \pi^+\pi^-\pi^0 n$.
We also suggest that the bottomed meson decay mode $B$$\to$$K\bar{K}^*\bar{D}^{(*)}$$\to$$KD_s^{(*)-}\pi\pi$ should be a promising process for the study of the ATS. In this process the peak structure corresponding to the ATS will be located far away from the normal threshold.

We also pointed out that the ATS contribution may mix with that produced by a genuine pole near threshold. Such ambiguities can be clarified by studies of the energy-dependence of the invariant mass spectrum. Different production processes can also provide additional information for the nature of the threshold enhancements. For some $XYZ$ particles, the presence of the ATS means that a combined study of the ATS mechanism and other dynamic processes are necessary. This should be crucial for our better understanding of those $XYZ$ threshold enhancements. Further experimental studies of the ATS at BES-III, Belle-II and LHCb would be extremely valuable for clarifying many existing puzzles.

\subsection*{Acknowledgments}

Useful discussions with F. K. Guo, G. Li and Q. Wang are acknowledged.
This work was supported, in part, by the Japan Society for the Promotion of Science under Contract No. P14324, the JSPS KAKENHI (Grant No. 25247036),
the National Natural Science Foundation of China (Grant No.
11425525), and the Sino-German CRC 110 ``Symmetries and
the Emergence of Structure in QCD" (NSFC Grant No. 11261130311).


\begin{thebibliography}{99}

\bibitem{Guo:2014iya}
F.~K.~Guo, C.~Hanhart, Q.~Wang and Q.~Zhao,
Phys.\ Rev.\ D {\bf 91}, no. 5, 051504 (2015)
[arXiv:1411.5584 [hep-ph]].


\bibitem{Peierls:1961zz}
R.~F.~Peierls,
Phys.\ Rev.\ Lett.\  {\bf 6}, 641 (1961).


\bibitem{Goebel:1964zz}
C.~Goebel,
Phys.\ Rev.\ Lett.\  {\bf 13}, 143 (1964).


\bibitem{hwa:1963aa}
R.~C.~Hwa,
Phys.\ Rev.\   {\bf 130}, 2580 (1963).

\bibitem{landshoff:1962aa}
P.~Landshoff and S.~Treiman,
Phys.\ Rev.\   {\bf 127}, 649 (1962).



\bibitem{Aitchison:1969tq}
I.~J.~R.~Aitchison and C.~Kacser,
Phys.\ Rev.\  {\bf 173}, 1700 (1968).


\bibitem{BESIII:2012aa}
M.~Ablikim {\it et al.}  [BESIII Collaboration],
Phys.\ Rev.\ Lett.\  {\bf 108}, 182001 (2012)
[arXiv:1201.2737 [hep-ex]].


\bibitem{Wu:2011yx}
J.~J.~Wu, X.~H.~Liu, Q.~Zhao and B.~S.~Zou,
Phys.\ Rev.\ Lett.\  {\bf 108}, 081803 (2012)
[arXiv:1108.3772 [hep-ph]].


\bibitem{Wu:2012pg}
X.~G.~Wu, J.~J.~Wu, Q.~Zhao and B.~S.~Zou,
Phys.\ Rev.\ D {\bf 87}, no. 1, 014023 (2013)
[arXiv:1211.2148 [hep-ph]].


\bibitem{Aceti:2012dj}
F.~Aceti, W.~H.~Liang, E.~Oset, J.~J.~Wu and B.~S.~Zou,
Phys.\ Rev.\ D {\bf 86}, 114007 (2012)
[arXiv:1209.6507 [hep-ph]].


\bibitem{Ketzer:2015tqa}
M.~Mikhasenko, B.~Ketzer and A.~Sarantsev,
Phys.\ Rev.\ D {\bf 91}, no. 9, 094015 (2015)
[arXiv:1501.07023 [hep-ph]].


\bibitem{wu-etal}
X. G. Wu, X. H. Liu, Q. Zhao and B. S. Zou, work to be submitted.

\bibitem{Ablikim:2013mio}
M.~Ablikim {\it et al.}  [BESIII Collaboration],
Phys.\ Rev.\ Lett.\  {\bf 110}, 252001 (2013)
[arXiv:1303.5949 [hep-ex]].


\bibitem{Ablikim:2013emm}
M.~Ablikim {\it et al.}  [BESIII Collaboration],
Phys.\ Rev.\ Lett.\  {\bf 112}, no. 13, 132001 (2014)
[arXiv:1308.2760 [hep-ex]].


\bibitem{Liu:2013dau}
Z.~Q.~Liu {\it et al.}  [Belle Collaboration],
Phys.\ Rev.\ Lett.\  {\bf 110}, 252002 (2013)
[arXiv:1304.0121 [hep-ex]].


\bibitem{Wang:2013cya}
Q.~Wang, C.~Hanhart and Q.~Zhao,
Phys.\ Rev.\ Lett.\  {\bf 111}, no. 13, 132003 (2013)
[arXiv:1303.6355 [hep-ph]].


\bibitem{Liu:2013vfa}
X.~H.~Liu and G.~Li,
Phys.\ Rev.\ D {\bf 88}, 014013 (2013)
[arXiv:1306.1384 [hep-ph]].


\bibitem{Pakhlov:2014qva}
P.~Pakhlov and T.~Uglov,
arXiv:1408.5295 [hep-ph].


\bibitem{Landau:1959fi}
L.~D.~Landau,
Nucl.\ Phys.\  {\bf 13}, 181 (1959).


\bibitem{bonnevay:1961aa}
G.~Bonnevay, I.~J.~R.~Aitchison and J.~S.~Dowker,
Nuovo\ Cim.\  {\bf 21}, 3569 (1961).



\bibitem{Eden:1966}
R.~J.~Eden, P.~V.~Landshoff, D.~I.~Olive and J.~C.~Polkinghorne,
{\it The Ananytic S-Matrix}, Cambridge University Press 1966.

\bibitem{Cutkosky:1960sp}
R.~E.~Cutkosky,
J.\ Math.\ Phys.\  {\bf 1}, 429 (1960).


\bibitem{barton:1961aa}
G.~Barton and C.~Kacser,
Nuovo\ Cim.\  {\bf 21}, 3161 (1961).

\bibitem{fronsdal:1964aa}
C.~Fronsdal and R.~E.~Norton,
J.\ Math.\ Phys.\ {\bf 5}, 100 (1964).

\bibitem{bronzan:1964aa}
J.~B.~Bronzan
Phys.\ Rev.\   {\bf 134}, B687 (1964).

\bibitem{norton:1964aa}
R.~E.~Norton
Phys.\ Rev.\   {\bf 135}, B1381 (1964).

\bibitem{Lucha:2006vc}
W.~Lucha, D.~Melikhov and S.~Simula,
Phys.\ Rev.\ D {\bf 75}, 016001 (2007)
[hep-ph/0610330].


\bibitem{coleman:1965aa}
S.~Coleman and R.~E.~Norton
Nuovo\ Cim.\  {\bf 38}, 438 (1965).


\bibitem{schmid:1967aa}
C.~Schmid
Phys.\ Rev.\   {\bf 154}, 1363 (1967).



\bibitem{Goebel:1982yb}
C.~J.~Goebel, S.~F.~Tuan and W.~A.~Simmons,
Phys.\ Rev.\ D {\bf 27}, 1069 (1983).


\bibitem{Anisovich:1995ab}
A.~V.~Anisovich and V.~V.~Anisovich,
Phys.\ Lett.\ B {\bf 345}, 321 (1995).

\bibitem{Bugg:2011jr}
  D.~V.~Bugg,
  Europhys.\ Lett.\  {\bf 96}, 11002 (2011)
  doi:10.1209/0295-5075/96/11002
  [arXiv:1105.5492 [hep-ph]].


\bibitem{Casalbuoni:1996pg}
R.~Casalbuoni, A.~Deandrea, N.~Di Bartolomeo, R.~Gatto, F.~Feruglio and G.~Nardulli,
Phys.\ Rept.\  {\bf 281}, 145 (1997)
[hep-ph/9605342].



\bibitem{Aaij:2014baa}
R.~Aaij {\it et al.} [LHCb Collaboration],
Phys.\ Rev.\ D {\bf 90}, no. 7, 072003 (2014)
[arXiv:1407.7712 [hep-ex]].

\bibitem{Agashe:2014kda}
K.~A.~Olive {\it et al.}  [Particle Data Group Collaboration],
Chin.\ Phys.\ C {\bf 38}, 090001 (2014).


\bibitem{Adolph:2015pws}
C.~Adolph {\it et al.}  [COMPASS Collaboration],
arXiv:1501.05732 [hep-ex].


\bibitem{Xiao:2013iha}
T.~Xiao, S.~Dobbs, A.~Tomaradze and K.~K.~Seth,
Phys.\ Lett.\ B {\bf 727}, 366 (2013)
[arXiv:1304.3036 [hep-ex]].


\bibitem{Wang:2013hga}
Q.~Wang, C.~Hanhart and Q.~Zhao,
Phys.\ Lett.\ B {\bf 725}, no. 1-3, 106 (2013)
[arXiv:1305.1997 [hep-ph]].


\bibitem{Cleven:2013mka}
M.~Cleven, Q.~Wang, F.~K.~Guo, C.~Hanhart, U.~G.~Meißner and Q.~Zhao,
Phys.\ Rev.\ D {\bf 90}, no. 7, 074039 (2014)
[arXiv:1310.2190 [hep-ph]].


\bibitem{Wang:2013kra}
Q.~Wang, M.~Cleven, F.~K.~Guo, C.~Hanhart, U.~G.~Meissner, X.~G.~Wu and Q.~Zhao,
Phys.\ Rev.\ D {\bf 89}, no. 3, 034001 (2014)
[arXiv:1309.4303 [hep-ph]].


\bibitem{Liu:2014spa}
X.~H.~Liu,
Phys.\ Rev.\ D {\bf 90}, no. 7, 074004 (2014)
[arXiv:1403.2818 [hep-ph]].


\bibitem{Wu:2013onz}
X.~G.~Wu, C.~Hanhart, Q.~Wang and Q.~Zhao,
Phys.\ Rev.\ D {\bf 89}, no. 5, 054038 (2014)
[arXiv:1312.5621 [hep-ph]].


\bibitem{Szczepaniak:2015eza}
A.~P.~Szczepaniak,
Phys.\ Lett.\ B {\bf 747}, 410 (2015)
[arXiv:1501.01691 [hep-ph]].

\end{thebibliography}
\end{document}